\documentclass[referee]{jfm2}
\usepackage{amsmath, amssymb}
\usepackage[dvips]{graphicx}
\usepackage{natbib}
\usepackage{bm}

\def\hf{{\frac{1}{2}}}
\begin{document}

\bibliographystyle{plainnat}
\title[Explicit radius for nearly parallel vortex filaments in equilibrium]{Explicit mean-field radius for nearly parallel vortex filaments in statistical equilibrium}
\author[T. D. Andersen and C. C. Lim]{Timothy D. Andersen \thanks{andert@rpi.edu}
 \and Chjan C. Lim \thanks{limc@rpi.edu}}

\affiliation{Mathematical Sciences, RPI, 110 8th St., Troy, NY, 12180}

\date{\today}

\begin{abstract}
Geophysical research has focused on flows, such as ocean currents, as two dimensional. Two
dimensional point or blob vortex models have the advantage of having a Hamiltonian, whereas 3D
vortex filament or tube systems do not necessarily have one, although they do have action functionals.
On the other hand, certain classes of 3D vortex models called nearly parallel vortex filament models do have a
Hamiltonian and are more accurate descriptions of geophysical and atmospheric flows than purely 2D
models, especially at smaller scales. In these ``quasi-2D'' models we replace 2D point vortices with vortex
filaments that are very straight and nearly parallel but have Brownian variations along their lengths due to local self-induction. When very straight, quasi-2D filaments are expected to have virtually the same planar density
distributions as 2D models. An open problem is when quasi-2D model statistics behave differently
than those of the related 2D system and how this difference is manifested.  In this paper we study the nearly parallel vortex filament
model of \cite{Klein:1995} in statistical equilibrium.  We are able to obtain a free-energy functional for the
system in a non-extensive thermodynamic limit that is a function of the mean square vortex position
$R^2$ and solve \emph{explicitly} for $R^2$.  Such an
explicit formula has never been obtained for a non-2D model.  We compare the results of our
formula to a 2-D formula of \cite{Lim:2005} and show qualitatively different behavior even when we disallow vortex braiding.  We further confirm our
results using Path Integral Monte Carlo (\cite{Ceperley:1995}) \emph{without} permutations and that the \cite{Klein:1995} model's
asymptotic assumptions \emph{are valid} for parameters where these deviations occur. 
\end{abstract}

\maketitle

\section{Introduction}

\begin{figure}
\begin{center}
\includegraphics[width = \textwidth]{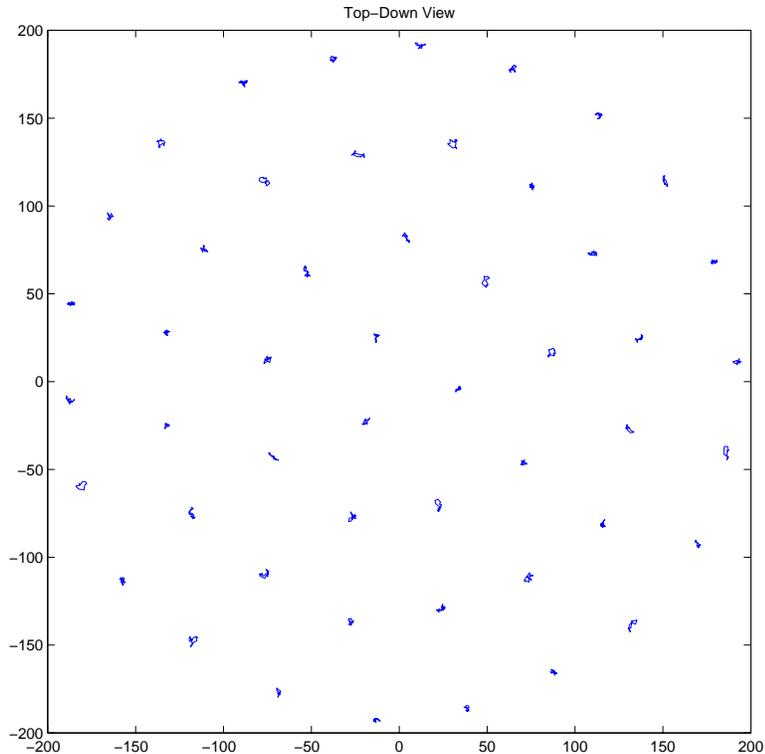}
\end{center}
\caption{Shown here in top-down projection, nearly parallel vortex filaments, at low density and high strength of interaction,
are well-ordered into a 2-D triangular lattice known as the Abrikosov lattice from type-II superconductors (\cite{Abrikosov:1957}).  This figure shows how the quasi-2D model is essentially
a 2-D model for these parameters.}
\label{fig:paths2d}
\end{figure}

In the study of geophysical turbulence, statistical equilibrium theories have focused mainly on 2-D models, which are known to have significant differences from the realistic 3-D.  Of particular interest in the 2-D regime is the vortex density distribution in the plane and how this density carries over to a quasi-2D regime.  In the strictly 2D setting, Lim and Assad have
variationally derived a low-temperature formula for the variance in vortex position, $R^2$, for the Onsager model in the Gibbs canonical ensemble with equal strength vortices in which
\begin{equation}
 R^2 = \frac{\beta N}{4\mu},
\label{eqn:rsq2D}
\end{equation} where $N$ is the number of filaments all with unity circulation, $\beta$ is the inverse temperature, and 
$\mu$ is the Lagrange multiplier for the angular momentum.  They have confirmed this formula and shown that the density is essentially uniform and axisymmetric using Monte Carlo simulations indicating that $R^2$ is the only significant moment at low-temperature (\cite{Lim:2005}).  While we assume that extremely straight filaments have a density profile
similar to that of the 2-D model and in turn a similar $R^2$ (Figure \ref{fig:paths2d}), we must also accept that at some magnitude of curviness the
$R^2$ for the nearly parallel model differs significantly from that of the fully parallel model.  
The entropy internal of each filament due to self-induction alone must at some point affect the profile.  What the difference in profiles is and for what parameters it is significant are open problems.

In this paper we address the problem of when $R^2$ of the 2-D model agrees with that of the nearly parallel vortex filament model of Klein, Majda, and Damodaran (KMD) (\cite{Klein:1995}), the simpliest
model of 3-D interacting filaments.  We solve explicitly for $R^2$ in the case of nearly parallel vortex filaments using a simple
mean-field approximation for the interaction potential and a spherical constraint.  We show that with current mathematical knowledge a spherical constraint is crucial to solving for $R^2$, making ours a novel application of the constraint
and the first analytical expression for $R^2$ in a non-2D model of vorticity.

In a previous paper (\cite{Andersen2:2006}) we solved for $R^2$ in a broken segment model.  Here we address the \emph{continuum} model and find that our mean-field energy functional is equivalent to that of a
quantum harmonic oscillator with a spherical constraint and a constant force.  Therefore, to solve for $R^2$, we apply a modified version of the spherical quantum methods of Hartman and Weichman (\cite{Hartman:1995}) to derive the free-energy of the system $f$ as a function of $R^2$ and minimize $f$ w.r.t. $R^2$, giving
\begin{align}
 R^2 = \frac{\beta^2\alpha N + \sqrt{\beta^4\alpha^2N^2 + 32\alpha\beta\mu}}{8\alpha\beta\mu},
\label{eqn:rsq3D}
\end{align} where $\alpha$ is the core-strength parameter and affects how curvy the lines are (\cite{Klein:1995}), $N$ is the number of filaments, and
$\mu$ and $\beta$ are the same as in Equation \ref{eqn:rsq2D}.  We show that equation \ref{eqn:rsq3D}, in the $\alpha\rightarrow\infty$ limit (making the lines perfectly straight), approaches equation \ref{eqn:rsq2D} (\S \ref{sec:meanfield}).

To confirm our quasi-2D formula and that we do not violate the KMD model's asymptotic assumptions (vortices very straight and far apart enough so they do not braid),
we perform Path Integral Monte Carlo (PIMC) (\cite{Ceperley:1995}) on the KMD system in the Gibbs canonical ensemble at different
 inverse temperatures, $\beta$.  We adapt PIMC to our model by eliminating the permutation
sampling used for quantum bosons (\cite{Ceperley:1995}) and type-II superconductors (\cite{Nordborg:1998}) and adapt it
to sample exactly the normally distributed, non-interacting filament case of self-induction \emph{plus} angular momentum so that the Monte Carlo's rejections depend only on interaction energy (\S \ref{sec:mcmodel}).  

For our chosen parameters, the
Monte Carlo shows that the model is valid down to a certain value of $\beta$ beyond which vortices begin to braid, and it shows that filament straightness assumptions hold for even smaller $\beta$.  Further, it confirms that $\beta$ values do exist where no braiding occurs \emph{and} our quasi-2D formula for
$R^2$ (Equation \ref{eqn:rsq3D}) gives a better prediction than Equation \ref{eqn:rsq2D}, a novel result (\S \ref{sec:mcresults}).

\section{Problem and Model}
\label{sec:model}
Geophysical vortex models are almost exclusively two dimensional because, on a planetary scale or
even the scale of several hundred or thousand kilometers, oceanic and atmospheric currents have
only a tiny vertical component.  Models like the 2-D Onsager N-Point-Vortex Gas model are nice
and simple to treat mathematically.

The Onsager Gas Model uses the Hamiltonian,
\begin{equation}
 H_N^{2D} = -\sum_{j>i}\lambda_i\lambda_j\log|\psi_i - \psi_j|^2,
\end{equation} where $\lambda_i$ and $\lambda_j$ are the circulation constants for point vortices $i$ and $j$ and $\psi_i$ and $\psi_j$ are their positions in the complex plane.  The Gibbs canonical ensemble,
\begin{equation}
 G_N = \frac{\exp{(-\beta H_N - \mu I_N)}}{Z_N},
\label{eqn:gibbs}
\end{equation} where
\begin{equation}
 Z_N = \int d\psi_1\cdots\int d\psi_N \exp{(-\beta H_N - \mu I_N)}
\label{eqn:partfn}
\end{equation} and 
\begin{equation}
 I_N^{2D} = \sum_i^N \lambda_i|\psi_i|^2,
\end{equation} represents the statistical distribution of the vortices in the gas
 with an inverse temperature $\beta$ and angular momentum chemical potential $\mu$ (\cite{Onsager:1949}).
From this model, Lim and Assad were able to derive variationally a low-temperature (large $\beta$) formula for
the second moment of the density $R^2$ (Equation \ref{eqn:rsq2D}) and describe the shape of the density profile as being nearly
uniform with a sharp density cutoff at $2R$ when $\lambda_i=\lambda_j\forall i,j$ (\cite{Lim:2005}).  Because the Onsager model is a 
Hamiltonian system such an approach is possible.

On the other hand, three dimensional models usually have no Hamiltonian representation and
models need to simulate the Biot-Savart interaction directly and obtain stationary states from
the Euler equations, a computationally intensive and analytically difficult (if not intractable) task.
Despite this difficulty, 3-D models are important to the study of realistic geophysics.

\begin{figure}
\begin{center}
\includegraphics[width = \textwidth]{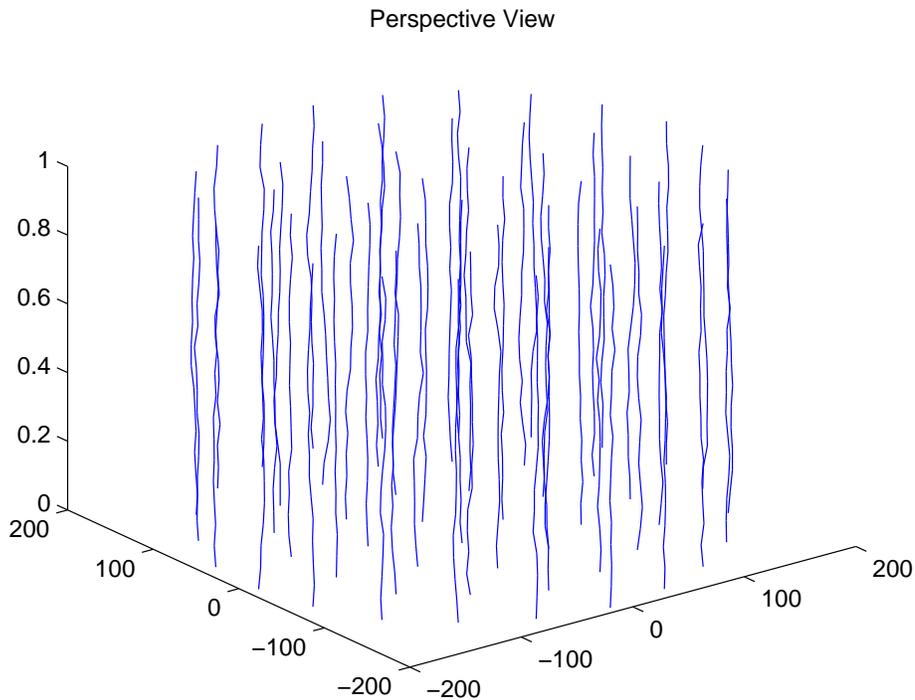}
\end{center}
\caption{Shown here in perspective projection, nearly parallel vortex filaments are quasi-2D.}
\label{fig:paths3d}
\end{figure}

At a certain scale vortex filaments become not quite 3-D but not quite 2-D either and
what we term a \emph{quasi}-2D model is applicable, one that takes into account the variation in the vorticity field in the vertical direction but does not allow for hairpins or loops or tangles in filaments (Figure \ref{fig:paths3d}).  

A restricted
quasi-2D model due to Klein, Majda, and Damodaran (\cite{Klein:1995}), derived from the Navier-Stokes
equations, represents vorticity as a bundle of $N$ filaments that are nearly parallel to the z-axis.
They have a Hamiltonian,
\begin{equation}
H_N = \alpha\int_0^Ld\sigma \sum_{k=1}^{N} \frac{1}{2}\left|\frac{\partial \psi_k(\sigma)}{\partial\sigma}
\right|^2 - \int_0^Ld\sigma\sum_{k=1}^N\sum_{i>k}^N \log|\psi_i(\sigma) - \psi_k(\sigma)|,
\label{eqn:Ham}
\end{equation} where $\psi_j(\sigma) = x_j(\sigma) + iy_j(\sigma)$ is the position of vortex $j$ at
position $\sigma$ along its length, the circulation constant is same for all vortices and set to $1$, and $\alpha$ is the core structure constant
 (\cite{Klein:1995}).  The
position in the complex plane, $\psi_j(\sigma)$, is assumed to be periodic in $\sigma$ with period $L$.  Self-stretching is neglected in the model although stretching due to interaction occurs.
Because this system does have a Hamiltonian, we can form a Gibbs distribution (Equation \ref{eqn:gibbs}) for it and study it
in statistical equilibrium with conservation of angular momentum,
\begin{equation}
 I_N = \sum_i^N \int_0^L d\sigma |\psi_i(\sigma)|^2.
\end{equation}  This allows us to study the density profile of the vortices.

For a large range of parameters, the probability distribution for this quasi-2D model shows very 
little difference from that of the fully 2-D model of Onsager because the variation in the z-axis
direction is too small to have any effect.  Therefore, the model begs the question of when the
density profile begins to behave differently, when do 3-D effects come into play, when does
this model succeed while the 2-D model fails.  

We hypothesize that there exists and set out the find parameters ($\alpha$,$\beta$,$\mu$) such that the most
important statistical moment, the second one $R^2$, differs in the two systems. (The mean is always the origin in this system so by default the second is the most significant moment.)  Furthermore,
we hypothesize that for at least part of this range of parameters the assumptions of the
quasi-2D model hold, where these assumptions state that
\begin{enumerate}
 \item for a given rise along a filament $\epsilon$, the variation in the complex plane is order $\epsilon^2$, where $\epsilon\ll 1$
 \item vortices do not form braids
 \item the core-size $\delta\ll\epsilon$, which we assume.
\end{enumerate}

We begin by deriving a mean-field, spherically constrained, explicit expression for $R^2_{3D}$, Equation \ref{eqn:rsq3D}, 
and we show that, in the $\alpha\rightarrow\infty$ limit, causing
the vortices to become perfectly straight and parallel, $R^2_{3D}\rightarrow R^2_{2D}$, that is that as
the quasi-2D model becomes fully 2D the expression for $R^2_{3D}$ becomes 2D as well (\S\ref{sec:meanfield}).  We then
verify our expression and the second hypothesis, that the assumptions are not broken, using Quantum Monte Carlo methods of Ceperley (\cite{Ceperley:1995}) (\S\ref{sec:mcresults}).

\section{Mean-field Theory}
\label{sec:meanfield}
The partition function for the quasi-2D system,
\begin{equation}
 Z_N = \int D\psi_1\cdots \int D\psi_N \exp\left(S_N\right),
\label{eqn:partfn2}
\end{equation} where $D\psi_k$ represents functional integration over all Feynman paths and
$S_N = -\beta H_N - \mu I_N$ is the action functional gives us the free energy for the most
probable macrostate:
\begin{equation}
 F = -\frac{1}{\beta}\log Z_N.
\end{equation}
 This free energy for the quasi-2D Hamiltonian (Equation \ref{eqn:Ham}) cannot
be found analytically with current mathematical knowledge.  In fact, the partition function for
the 2-D point-vortex system cannot be found for a finite number of vortices.  The reason for
the difficulty comes from the interaction term (second term in Equation \ref{eqn:Ham}).  While the other terms, the self-induction (first term in Eq. \ref{eqn:Ham})
and the conservation of angular momentum term, are quadratic and yield a normally distributed
Gibbs distribution that we can functionally integrate, the logarithmic term must be approximated.

Although Lions and Majda have made such an approximation and rigorously derived a mean-field PDE for the
probability distribution of the vortices in the complex plane (\cite{Lions:2000}), their
PDE takes the form of a non-linear Schroedinger equation that is not analytically solvable, again
because of the interaction term.  While a discussion of their
derivation is beyond the scope of this paper, their existence proof justifies our applying a much simpler mean-field approximation to interaction.  

For
very straight filaments that are well-ordered, we can take the mean-field action to be
\begin{equation}
 S_N^{mf} = -\int_0^L \quad d\sigma \left[\sum_{k=1}^{N} \frac{\beta\alpha}{2}\left|\frac{\partial \psi_k(\sigma)}{\partial\sigma}
\right|^2 + \frac{N\beta}{4}\sum_{k=1}^{N}\log |\psi_k(\sigma)|^2 - \mu\sum_{k=1}^N|\psi_k(\sigma)|^2\right],
\label{eqn:ActionApprox}
\end{equation} where the second term in the integrand is now mean-field and the other two are the same as before.  This mean-field approximation derives from the
work of Lim and Assad for point-vortices (\cite{Lim:2005}).  Intuitively it says that on average the distance between vortices is also the distance of a vortex from the center.  Based on Monte Carlo simulations, this approximation works extremely well even for high levels of filament variation.

Before we begin to calculate the partition function, we must deal with another problem:
we still cannot integrate Equation \ref{eqn:partfn2} using this action because the interaction term (second term in \ref{eqn:ActionApprox}) is still a function of $\psi$ and still inside
the integral, so we make another approximation, adding a microcanonical constraint on the
angular momentum,
\begin{equation}
 \delta\left(\int_0^L d\sigma \left[|\psi(\sigma)|^2 - R^2\right]\right),
\end{equation} that has integral representation,
\begin{equation}
 \delta\left(\int_0^L d\sigma \left[|\psi(\sigma)|^2 - R^2\right]\right) = \int_{-\infty}^{\infty}\frac{d\tau}{2\pi} \exp \int_0^L -i\tau\left[|\psi(\sigma)|^2 - R^2\right].
\label{eqn:sphere}
\end{equation}  This approximation only applies when
$\mu$ is large and fluctuations in the angular momentum (and consequently $R^2$) are fairly small, but
it allows us to remove the interaction term from the functional integral \ref{eqn:partfn2}. 

Now the spherical-mean-field partition function looks like
\begin{equation}
 Z_N^{smf} = \int D\psi_1\cdots \int D\psi_N \exp\left(S_N\right)\int_{-\infty}^{\infty}\frac{d\tau}{2\pi} \exp \int_0^L d\sigma -i\tau\left[|\psi(\sigma)|^2 - R^2\right].
\end{equation}  \footnote{The Hartman and Weichman paper (\cite{Hartman:1995}) confines the wavefunction to the surface of a sphere
using an infinite number of Dirac delta functions and a functional integral over $\tau$, which
here would be like confining $\psi(\sigma)$ to a cylinder of radius $R$.  We do not apply their method
in this way but confine the \emph{mean} of $\psi$, $L^{-1}\int_0^L d\sigma\psi(\sigma)$, to that cylinder with
only one Dirac delta.} and, since the exponents are all positive definite, we can interchange the
integrals and combine exponents,
\begin{equation}
 Z_N^{smf} = \int_{-\infty}^{\infty}\frac{d\tau}{2\pi}\int D\psi_1\cdots \int D\psi_N \exp\left(S_N^{smf}\right).
\end{equation} where the action functional is
\begin{align}
 S_N^{smf} = \sum_{k=1}^N S_k
\end{align} and
\begin{align}
 S_k = \left[\beta LN\log(R^2)/4 - \hf\int_0^L d\sigma \alpha\beta\left|\frac{\partial \psi_k(\sigma)}{\partial\sigma}\right|^2 + (i\tau + 2\mu)|\psi_k(\sigma)|^2 - i R^2\tau(\sigma)\right]
\end{align} is the single filament action.  Because $\psi_k$ are statistically independent for all $k$, we
let $S=S_k\forall k$ and $S_N^{smf} = NS$, which makes the partition function,
\begin{equation}
 Z_N^{smf} = \int_{-\infty}^{\infty}\frac{d\tau}{2\pi}\left\{\int D\psi \exp S\right\}^N,
\end{equation}.

We need to have the partition function in the form $Z_N^{smf} = \int d\tau e^{-Nf}$ to use steepest descent (Appendix \ref{sec:spherical}), and so we define the non-dimensional free energy as a function of $i\tau$, $f[i\tau] = \beta F$,
\begin{equation}
 f[i\tau] = -\log\left[\int D\psi\exp\left(S\right)\right].
\label{eqn:fintegral}
\end{equation} which allows us to re-write the partition function as
\begin{equation}
 Z_N^{smf} = \int_{-\infty}^{\infty}\frac{d\tau}{2\pi}\exp\left(-Nf[i\tau]\right).
\end{equation}  This form of the partition function we can solve with steepest descent methods,
but first we need a formula for $f$.

We can evaluate $f[i\tau]$, the energy of a 2-D quantum harmonic 
oscillator with a constant force, with Green's function methods but do not go into it in this paper (\cite{Brown:1992}).
Let us make a change of variables $\lambda = i\tau + 2\mu$ and the non-extensive scaling $\beta'=\beta N$ and
$\alpha' = \alpha/N$, which will become necessary when we take the limit $N\rightarrow\infty$.  After evaluating the integral, Equation \ref{eqn:fintegral}, (see Appendix \ref{sec:freenergy}), the free-energy reads
\begin{equation}
 f[\lambda] = L\mu - \hf L\lambda R^2 - \beta'L \log(R^2)/4 - \ln\frac{e^{-\omega L}}{\left(e^{-\omega L} - 1\right)^2},
\label{eqn:foflambdaAndR}
\end{equation} 
where $\omega = \sqrt{\lambda/(\alpha'\beta')}$ is the harmonic oscillator frequency. 

Now that we have a formula for $f$ we can apply the saddle point or steepest descent method.  (For discussion of this method see Appendix \ref{sec:spherical} as
well as the original paper of Berlin and Kac (\cite{Berlin:1952}).)  The intuition is that, as
$N\rightarrow\infty$ in the partition function, only the minimum energy will contribute to the integral, i.e. at infinite $N$, the exponential behaves like a Dirac delta function, so
\begin{equation}
 f_\infty = \lim_{N\rightarrow\infty}-\frac{1}{N}\ln Z_N^{smf} = f[\eta],
\end{equation} where $\eta$ is such that $\partial f[\lambda]/\partial \lambda|_\eta = 0$ (\cite{Hartman:1995},\cite{Berlin:1952}).

First we can make a simplification by ridding Equation \ref{eqn:foflambdaAndR} of $R^2$.  We know that $R^2$ will minimize $f$ and so $\partial f/\partial R^2 = 0$.
Since
\begin{equation}
 \frac{\partial f}{\partial R^2} = (\mu - \lambda/2)L - L\beta'/(4R^2),
\end{equation}
\begin{equation}
 R^2 = \frac{\beta'}{4(\mu - \lambda/2)}.
\label{eqn:rsq3d2}
\end{equation}

Substituting the left side of \ref{eqn:rsq3d2} for $R^2$ in Equation \ref{eqn:foflambdaAndR}, we get
\begin{equation}
 f[\lambda] = \beta' L/4 + \sqrt{\frac{\lambda}{\alpha'\beta'}}L + 2\log\left|\exp\left({-\sqrt{\frac{\lambda}{\alpha'\beta'}}L}\right) - 1\right| - \frac{\beta' L}{4}\log\frac{\beta'}{4(\mu - \lambda/2)}.
\label{eqn:foflambda}
\end{equation}

We could take the derivative of Equation \ref{eqn:foflambda} and set it equal to zero to obtain
$\eta$.  However, doing so yields a transcendental equation that needs to be solved numerically.  Since our goal is to obtain an explicit formula, we choose to study the system as $L\rightarrow\infty$.  In fact such an approach is justified by the assumptions of
the model that $L$ have larger order than the rest of the system's dimensions.  (If this were a quantum system, this procedure would be equilvalent to finding the energy of the ground state.  Hence, we call this energy $f_{grnd}$.)  Taking the limit on Equation \ref{eqn:foflambda} yields the free energy per unit length in which $\eta$ can be solved for
\begin{equation}
 f_{grnd}[\eta] = \frac{\beta'}{4} + \sqrt{\eta/(\alpha'\beta')} - \frac{\beta'}{4}\log\left(\frac{\beta'}{4(\mu - \eta/2)}\right),
\label{eqn:groundFreeEnergy}
\end{equation} where
\begin{equation}
 \eta = 2\mu - \frac{1}{8}\beta'(-\beta'^2\alpha' \pm \sqrt{\beta'^4\alpha'^2 + 32\alpha'\beta'\mu}),
\end{equation} of which we take
\begin{equation}
\eta = 2\mu - \frac{1}{8}\beta'(-\beta'^2\alpha' + \sqrt{\beta'^4\alpha'^2 + 32\alpha'\beta'\mu})
\end{equation} as giving physical results.

With $\eta$ explicit, we can give a full formula for $R^2$,
\begin{align}
 R^2 &= \frac{4}{-\beta'^2\alpha' + \sqrt{\beta'^4\alpha'^2 + 32\alpha'\beta'\mu}}\nonumber\\
&= \frac{\beta'^2\alpha' + \sqrt{\beta'^4\alpha'^2 + 32\alpha'\beta'\mu}}{8\alpha'\beta'\mu}.
\label{eqn:rsq3d3}
\end{align}

Through several approximations, we have obtained an explicit formula for the free energy of the
system and $R^2$.  Before we move on to the numerical verification, we can ask a few things about
it.  One question is how \ref{eqn:rsq3d3} compares to the 
formula that Lim and Assad derived variationally for point vortices, Equation \ref{eqn:rsq2D} (\cite{Lim:2005}).  We repeat that formula here:
\begin{equation}
 R^2_{2D} = \frac{N\beta}{4\mu}.
\label{eqn:rsq2d2}
\end{equation}

Because the quasi-2D system has only the self-induction term to differentiate its behavior from that of
 the Onsager Vortex Gas, if we make the lines perfectly straight through some limit, then the statistics of the quasi-2D system should be identical to that of the 2-D gas.  In the quasi-2D system $\alpha'$ controls the straightness of the lines.  The larger $\alpha'$ the more
inclined the lines are to be straight.  Therefore, if we take $\alpha'\rightarrow\infty$ limit on
Equation \ref{eqn:rsq3d3}, we should get the 2-D formula \ref{eqn:rsq2d2}:
\begin{align*}
 R^2(\alpha'\rightarrow\infty) &= \lim_{\alpha'\rightarrow\infty} \frac{4}{-\beta'^2\alpha' + \sqrt{\beta'^4\alpha'^2 + 32\alpha'\beta'\mu}} \\&=
\lim_{\gamma\rightarrow 0} \frac{4\gamma}{-\beta'^2 + \sqrt{\beta'^4 + 32\gamma\beta'\mu}},
\end{align*} where $\gamma=\alpha'^{-1}$.  We use L'Hopit\^al's rule taking derivatives of the
top and bottom
\begin{align}
 R^2(\alpha'\rightarrow\infty) &= \lim_{\gamma\rightarrow 0} \frac{4}{\hf(\beta'^4 + 32\gamma\beta'\mu)^{-\hf}32\beta'\mu}\nonumber\\
&= \frac{4\beta'^2}{16\beta'\mu}
= \frac{\beta'}{4\mu}
= \frac{N\beta}{4\mu}
\end{align} which agrees with Equation \ref{eqn:rsq2d2}.  That these equations agree indicates that,
like the point vortex equation, \ref{eqn:rsq3d3} is a low temperature formula but can predict behavior for all levels of \emph{internal} filament fluctuation.

There are qualitative differences in the two equations as well since they are different for finite $\alpha'$.  The 2-D equation, \ref{eqn:rsq2d2}, is linear in $\beta$, but Equation \ref{eqn:rsq3d3} is not.
In fact, a simple calculation (not shown here) tells us that, while \ref{eqn:rsq2d2} decreases linearly with
decreasing $\beta'$, Equation \ref{eqn:rsq3d3} begins to increase at a $\beta'=\beta'_0$ where
\begin{equation}
 \beta'^3_0 = \frac{4\mu}{\alpha'}.
\end{equation}  Therefore, for changing $\beta$, the 2-D Equation \ref{eqn:rsq2d2} is a straight
line, and the quasi-2D Equation \ref{eqn:rsq3d3} has a ``v''-shape, decreasing with $\beta'$ to
$\beta'_0$ and then increasing again.
Since the free energy in Equation \ref{eqn:groundFreeEnergy} is smooth for
all $\beta>0$ provided $\alpha>0,\mu>0$, this ``v''-shape does not indicate a phase transition but rather that at some $\beta$ the internal filament fluctuations overtake the conflict between the
angular momentum and interaction terms as the major contributor to the value of $R^2$.

We can also measure quantitative differences between \ref{eqn:rsq2d2} and
\ref{eqn:rsq3d3}.  We calculate the error as a function of the parameters:
\begin{equation}
 E = \frac{R^2 - R^2_{2D}}{R^2_{2D}},
\end{equation} which, after some simple algebra, gives the $\beta$ value at which a particular
error appears,
\begin{equation}
 \beta'^3 = \frac{8\mu}{\alpha' E(E + 1)}.
\label{eqn:error}
\end{equation}  This equation is the most useful for numerical verification because it
tells us exactly which parameters give how much error $E$, and error is what we aim to detect since
that will tell us when 3-D effects become noticeable.  In the next section we address whether we can detect the error in the Monte Carlo and without violating the assumptions of the model.

\section{Monte Carlo Comparison}
\label{sec:mcresults}
We apply Monte Carlo in this paper to the original quasi-2D model with Hamiltonian \ref{eqn:Ham} to verify two hypotheses:
\begin{enumerate}
 \item that the 3-D effects, namely the Equation \ref{eqn:rsq3d3}, predicted in
the mean-field are correct
 \item that these effects can be considered physical in the sense that
the model's asymptotic assumptions of straightness and non-braiding are not violated.
\end{enumerate}

We are caught between three requirements: to make the vortices wavy enough to show the effects we
predict but not so wavy as to violate either the assumptions of the model or the assumptions we
made in our derivation in \S\ref{sec:meanfield}.  The assumptions of the model must be violated at some small $\beta$ because decreasing $\beta$ increases fluctuation \emph{and} vortex density, meaning that vortices will begin to braid and ultimately become too wavy to be considered asymptotically straight.  (There are no other constraints on the model besides the filaments having very
small core-size, which we assume.)  Furthermore, because our assumptions make Equation \ref{eqn:rsq3d3} a low-temperature prediction, we need entropy--save that of internal fluctuations--to be small enough to neglect.
  In this section, we show not only that such a regime exists but that the mean-field equation, Equation \ref{eqn:rsq3d3}, predicts the mean square vortex position in the Monte Carlo, which we call $R^2_{MC}$, to better accuracy than the 2-D equation.

Research on flux-lines in type-II superconductors has yielded a close correspondence
between the behavior of vortex filaments in 3-space and paths of quantum bosons in (2+1)-D (2-space in
imaginary time) (\cite{Nordborg:1998},\cite{Sen:2001}).  This work is not related to ours fundamentally
because type-II superconductor flux-lines do not have the same boundary conditions.  They use
periodic boundaries in all directions with an interaction cut-off distance while we use no boundary conditions and no cut-off.  Besides the boundaries, they also
allow flux-lines to permute like bosons, switching the top endpoints, which we do not allow for our vortices because it would create unacceptable tangling.  However, despite the boundary differences, the
London free-energy functional for interacting flux-lines is closely related to our Hamiltonian
\ref{eqn:Ham}, and so we can apply Path Integral Monte Carlo (PIMC) in the same way as it has been 
applied to flux-lines.  (For a discussion of PIMC and how we apply it see Appendix \ref{sec:mcmodel}.)

We simulated a collection of $N=20$ vortices each with a piecewise linear representation with $M=1024$ segments and ran the system to equilibration, determined by the settling of the mean and variance of the total
energy.  We ran the system for 20 logarithmically spaced values of $\beta$ between $0.001$ and $1$ plus two points, $10$ and $100$.  We set $\alpha=10^7$, $\mu=2000$, and $L=10$.  We calculate several arithmetic averages: the mean square vortex position,
\begin{equation}
 R^2_{MC} = (MN)^{-1}\sum_{i=1}^N\sum_{k=1}^M|\psi_i(k)|^2,
\label{eqn:rsqMC}
\end{equation} where $k$ is the segment index corresponding to discrete values of $\sigma$, the mean square amplitude,
\begin{equation}
 A^2 = (MN)^{-1}\sum_{i=1}^N\sum_{k=1}^M|\psi_i(k)-\psi_i(1)|^2,
\label{eqn:amp}
\end{equation} which measures how wide the vortices are in top-down projection, the mean
square amplitude per segment,
\begin{equation}
 a^2 = (MN)^{-1}\sum_{i=1}^N\sum_{k=1}^M|\psi_i(k)-\psi_i(k+1)|^2, 
\label{eqn:aps}
\end{equation} where $\psi_i(M+1)=\psi_i(1)$, and the mean square nearest neighbor distance,
\begin{equation}
 d^2 = N^{-1}\sum_{i=1}^N\min_{j,k} |\psi_i(k)-\psi_j(k)|^2.
\label{eqn:nnd}
\end{equation}

\begin{figure}
\begin{center}
\includegraphics[width = \textwidth]{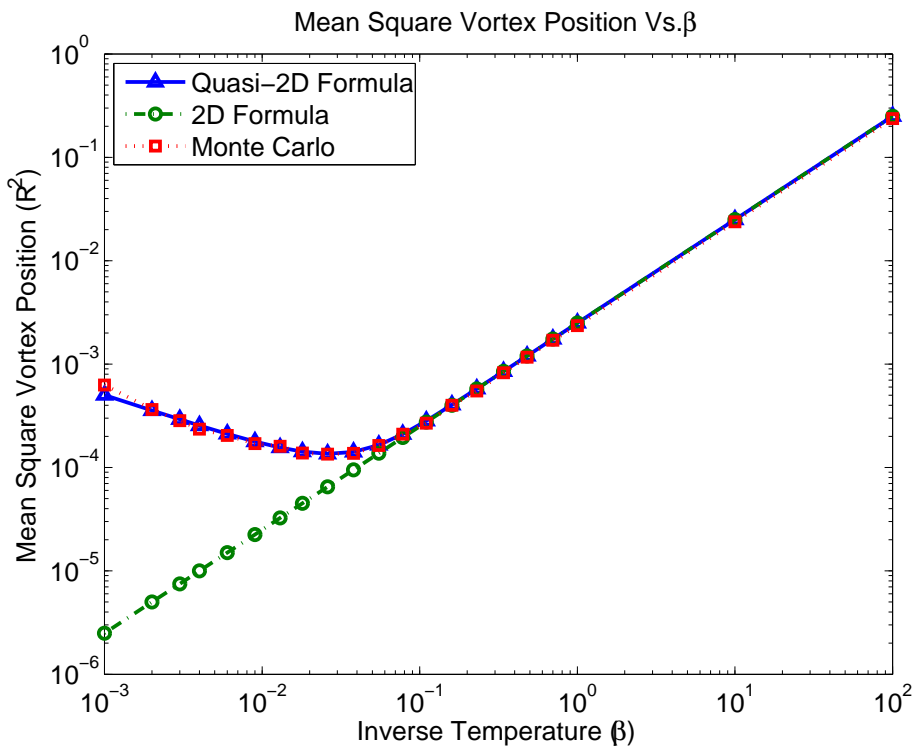}
\end{center}
\caption{The mean square vortex position, defined in Equation \ref{eqn:rsqMC}, compared with
Equations \ref{eqn:rsq3d3} and \ref{eqn:rsq2D} shows how 3-D effects come into play around
$\beta=0.16$.  That the 2D formula continues to decrease while the Monte Carlo and the quasi-2D
formula curve upwards with decreasing $\beta$ suggests that the internal variations of the
vortex lines have a significant effect on the probability distribution of vortices.}
\label{fig:mf2mcComp}
\end{figure}

Measures of Equation \ref{eqn:rsqMC} correspond well to Equation \ref{eqn:rsq3d3} in Figure 
\ref{fig:mf2mcComp} whereas Equation \ref{eqn:rsq2d2} continues to decline when the others curve
with decreasing $\beta$ values, suggesting that the 3-D effects are not only real in the Monte Carlo
but that the mean-field is a good approximation with these parameters.

\begin{figure}
\begin{center}
\includegraphics[width = \textwidth]{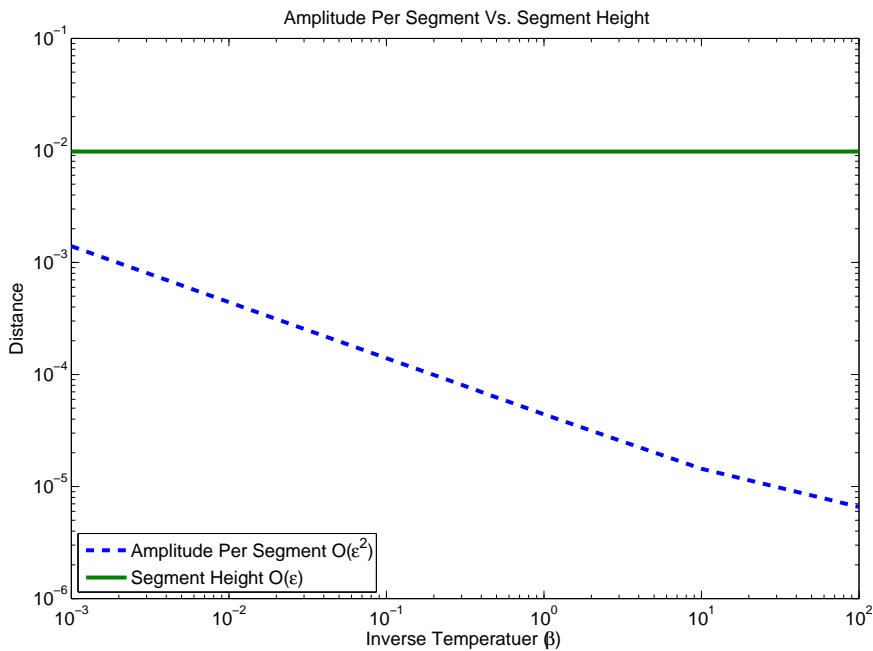}
\end{center}
\caption{This figure shows that the mean amplitude per segment (Equation \ref{eqn:aps}), meaning the distance in the complex plane between adjacent
points on the same piecewise-linear filament, is much less than the segment height $L/M = 10/1024$, indicating that
straightness constraints hold for all $\beta$ values.}
\label{fig:check1}
\end{figure}

\begin{figure}
\begin{center}
\includegraphics[width = \textwidth]{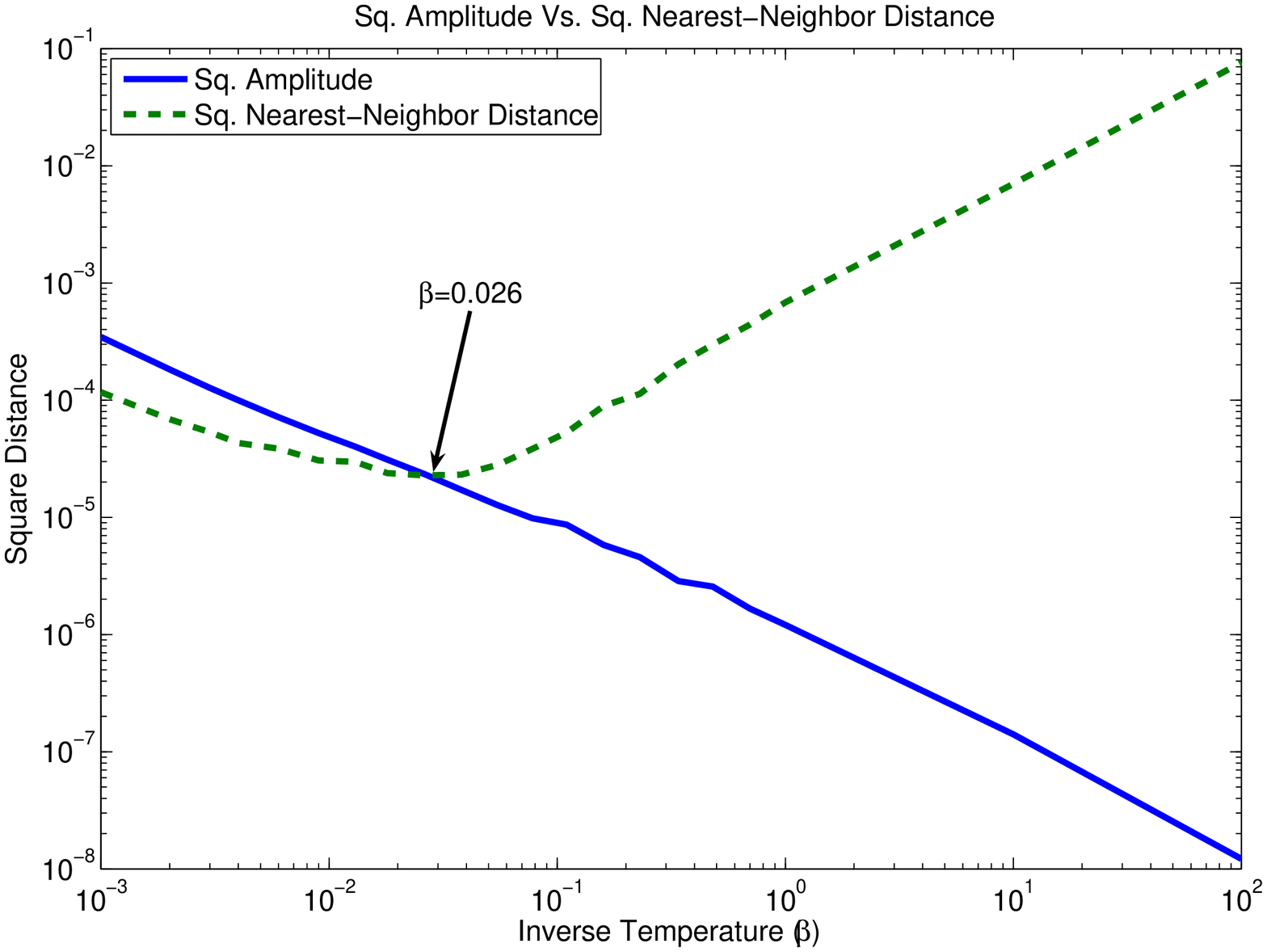}
\end{center}
\caption{The mean square amplitude (Equation \ref{eqn:amp}) becomes greater than the mean square
nearest-neighbor distance (Equation \ref{eqn:nnd}) at about $\beta=0.026$ indicating that
vortices begin to braid at this point.  Because braiding statistics are not covered in our
quasi-2D model, we cannot guarantee that the results for $\beta<0.038$ are physical.  At this point
braiding statistics are the subject of future research.}
\label{fig:check2}
\end{figure}

In order to be considered straight enough, we need
\begin{equation}
a \ll \frac{L}{M} = \frac{10}{1024}.
\end{equation}  Non-braiding requires
that $d^2 > A^2$.  According to Figure \ref{fig:check2}, braiding occurs around $\beta<0.038$ and straightness holds for all $\beta$ values, shown in Figure \ref{fig:check1}.  We do not need to show that these conditions hold for every instance in the Monte Carlo, 
only close to the average, because small probability events have little effect on the statistics.
We would also like to point out that
although we cannot guarantee that the statistical results for $\beta\in(0.002,0.038)$ are valid, we cannot dismiss them as being unphysical.  Whether the statistics for $\beta<0.038$ are physical is a subject for future research.  

\begin{figure}
\begin{center}
\includegraphics[width = \textwidth]{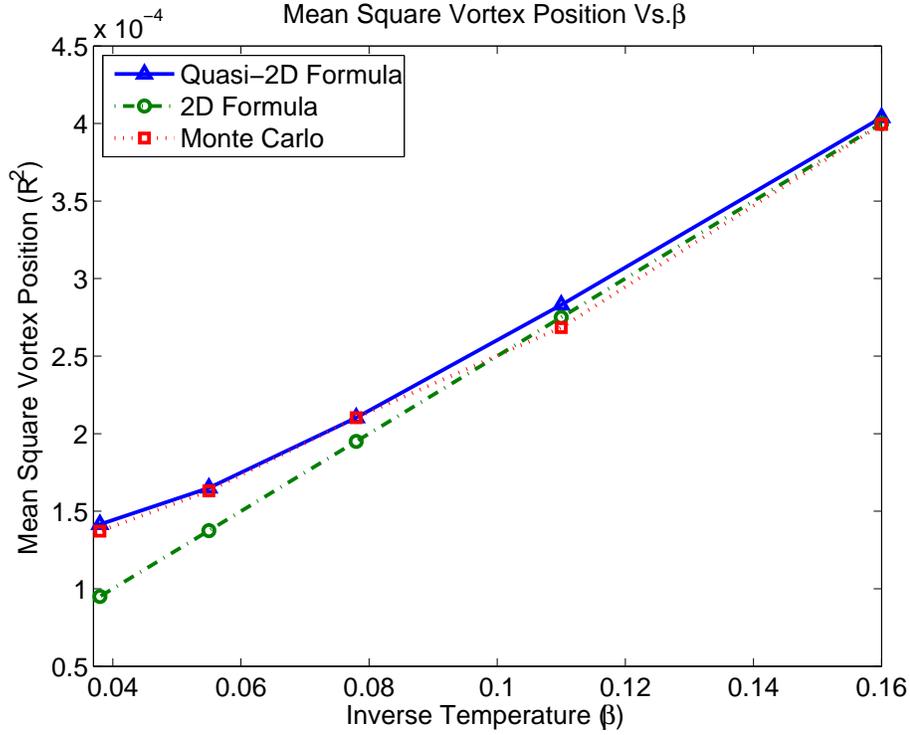}
\end{center}
\caption{The mean square vortex position, defined in Equation \ref{eqn:rsqMC}, compared with
Equations \ref{eqn:rsq3d3} and \ref{eqn:rsq2D} in close-up view between $\beta=0.16$ and $\beta=0.038$
where we can guarantee validity.}
\label{fig:mf2mcComp2}
\end{figure}

In Figure
\ref{fig:mf2mcComp2} we show a close-up of the most relevant $\beta$ values in the range $(0.038,0.16)$
 where there is greater correspondence between the Monte Carlo results and the quasi-2D formula
in Equation \ref{eqn:rsq3d3} than with Equation \ref{eqn:rsq2D}.  These results are most likely to
make a good physical prediction.

\section{Related Work}
\label{sec:related}
As mentioned in the previous section, simulations of flux lines in type-II superconductors using the PIMC method have been done, generating the Abrikosov lattice (\cite{Nordborg:1998},\cite{Sen:2001}).  However, the superconductor model has periodic boundary conditions in the xy-plane, is a different problem altogether, and is not applicable to trapped fluids.  No Monte Carlo studies of the model of \cite{Klein:1995} have been done to date
and dynamical simulations have been confined to a handful of vortices.  \cite{Kevlahan:2005} added a white noise term to the KMD Hamiltonian, Equation \ref{eqn:Ham}, to study vortex reconnection in
comparison to direct Navier-Stokes, but he confined his simulations to two vortices.  Direct
Navier-Stokes simulations of a large number of vortices are beyond our computational capacities.

\cite{Tsubota:2003} has done some excellent simulations of vortex tangles in He-4 with rotation, boundary walls, and \emph{ad hoc} vortex reconnections to study disorder in rotating superfluid turbulence.  Because vortex tangles are extremely curved, they applied the full Biot-Savart
law to calculate the motion of the filaments in time.  Their study did not include any sort of
comparison to 2-D models because for most of the simulation vortices were far too tangled.  The
inclusion of rigid boundary walls, although correct for the study of He-4, also makes the results 
only tangentially applicable to the KMD system we use.

Our use of the spherical model is recent and has also been applied to the statistical mechanics of macroscopic fluid flows in order to obtain exact solutions for quasi-2D turbulence (\cite{Lim1:2006}, \cite{Lim2:2006}).

Other related work on the statistical mechanics of turbulence in 3-D vortex lines can be found in \cite{Flandoli:2002} and \cite{Berdichevsky:1998} in addition to \cite{Lions:2000}.

\section{Conclusion}
\label{sec:conclusion}
We have developed an explicit mean-field formula for the most significant statistical moment for
the quasi-2D model of nearly parallel vortex filaments and shown that in Monte Carlo simulations
this formula agrees well while the related 2-D formula fails at higher temperatures.  We have also
shown that our predictions do not violate the model's asymptotic assumptions for a range of
inverse temperatures.  Therefore, we conclude that these results are likely physical.  We have
avoided braided vortices.  However, the model will admit braiding as long as vortices are
sufficiently straight and far apart enough.  In future we will address the problem of vortex
reconnections which can change the qualitative behavior of the system for some parameters.

\appendix
\section{Path Integral Monte Carlo method}
\label{sec:mcmodel}
Path Integral Monte Carlo methods emerged from the path integral formulation invented by Dirac 
that Richard Feynman later expanded (\cite{Zee:2003}), in which particles are conceived to follow
all paths through space.  One of Feynman's great contributions to
the quantum many-body problem was the mapping of path integrals onto a classical system of interacting
``polymers'' (\cite{Feynman:1948}).  D. M. Ceperley used Feynman's convenient piecewise linear formulation to develop his PIMC method which 
he successfully applied to He-4, generating the well-known lambda transition for the first time
in a microscopic particle simulation (\cite{Ceperley:1995}). Because it describes a system
of interacting polymers, PIMC applies to classical systems that have a ``polymer''-type
description like nearly parallel vortex filaments.  

PIMC has several advantages.  It is a
\emph{continuum} Monte Carlo algorithm, relying on no spatial lattice.  Only time (length in the z-direction
in the case of vortex filaments) is discretized, and the algorithm makes no assumptions about types of phase transitions or trial wavefunctions.

For our simulations we assume that the filaments are divided into
an equal number of segments of equal length.  This discretization leads to the Hamiltonian,

\begin{align}
H_N(M) = H_N^{self}(M) + H_N^{int}(M)
\end{align} where
\begin{equation}
 H_N^{self} = \alpha\sum_{j=1}^{M}\sum_{k=1}^{N} \frac{1}{2}\frac{|\psi_k(j+1) -
\psi_k(j)|^2}{\delta}
\end{equation} and
\begin{equation}
 H_N^{int} = - \sum_{j=1}^{M}\sum_{k=1}^N\sum_{i>k}^N\delta\log|\psi_i(j) - \psi_k(j)|,
\end{equation}
and angular momentum
\begin{equation}
I_N = \sum_{j=1}^{M}\sum_{k=1}^N \delta|\psi_k(j)|^2
,\end{equation} where $\delta$ is the length of each segment, $M$ is the number of segments,
and $\psi_k(j) = x_k(j) + iy_k(j)$ is the position of the point at which two segments meet (called in
PIMC a ``bead'') in the complex plane.

The probability distribution for vortex filaments,
\begin{equation}
G_N(M) = \frac{\exp{\left(-\beta H_N(M) - \mu I_N(M)\right)}}{Z_N(M)}
,\end{equation} where
\begin{equation}
Z_N(M) = \sum_{all paths} G_N(M),
\end{equation} is the Gibbs canonical distribution.  Our Monte Carlo simulations sample from this distribution.

The Monte Carlo simulation begins with a random distribution of filament end-points in a square of side
10, and there are two possible moves that the algorithm chooses at random.  The first is to move
a filament's end-points.  A filament is chosen at random, and its end-points moved a uniform random distance.  Then the energy of this new state, $s'$, is calculated and retained with probability
\begin{equation}
 A(s\rightarrow s') = \min\left\{1,\exp\left(-\beta [H_N^{int}(s')-H_N^{int}(s)] - \mu [I_N(s')-I_N(s)]\right)\right\},
\end{equation}  where $s$ is the previous state.  (Self-induction, $H_N^{self}$, is unchanged for this type of move since it is internal to each filament.)  The second move keeps end-points stationary and, following the bisection method of Ceperley, grows
a new internal configuration for a randomly chosen filament (\cite{Ceperley:1995}).  Both the
self-induction and the trapping potential are harmonic, so the Gibbs canonical distribution without interaction can be sampled directly as a Gaussian distribution.  Therefore, in this move the
configuration is generated by first sampling a free vortex filament, and
then accepting the new state with probability
\begin{equation}
 A(s\rightarrow s') = \min\left\{1, \exp\left[-\beta (H_N^{int}(s') - H_N^{int}(s))\right]\right\}.
\end{equation}

Our stopping criteria is graphical in that we ensure that the cumulative arithmetic mean of the energy settles to
a constant.  Typically, we run for 10 million moves or 50,000 sweeps for 200 vortices.  Afterwards, we
collect data from about 200,000 moves to generate statistical information.

\section{Spherical Model and the Saddle Point Method}
\label{sec:spherical}
The spherical model was first proposed in a seminal paper of Berlin and Kac (\cite{Berlin:1952}), in
which they were able to solve for the partition function of an Ising model given that the site spins
satisfied a spherical constraint, meaning that the squares of the spins all added up to a fixed
number.  The method relies on what is known as the saddle point or steepest descent approximation
method which is exact only for an infinite number of lattice sites.

In general the steepest descent or saddle-point approximation
applies to integrals of the form
\begin{equation}
 \int_a^b e^{-Nf(x)} dx,
\end{equation} where $f(x)$ is a twice-differentiable function,
$N$ is large, and $a$ and $b$ may be infinite.  A special case, called Laplace's method, concerns real-valued $f(x)$ with
a finite minimum value.

The intuition is that if $x_0$ is a point such that $f(x_0)< f(x)\forall x\neq x_0$, i.e. it is a global minimum, then, if we multiply
$f(x_0)$ by a number $N$, $Nf(x) - Nf(x_0)$ will be larger than
just $f(x) - f(x_0)$ for any $x\neq x_0$.  If $N\rightarrow\infty$ then
the gap is infinite.  For such large $N$, the only significant
contribution to the integral comes from the value of the integrand
at $x_0$.  Therefore,
\begin{equation}
 \lim_{N\rightarrow\infty} \left[\int_a^b e^{-Nf(x)}\right]^{1/N} dx = e^{-f(x_0)},
\end{equation} or
\begin{equation}
 \lim_{N\rightarrow\infty} -\frac{1}{N}\log\int_a^b e^{-Nf(x)} dx = f(x_0),
\end{equation} (\cite{Berlin:1952},\cite{Hartman:1995}).  A proof is easily obtained using
a Taylor expansion of $f(x)$ about $x_0$ to quadratic degree.

\section{Evaluating the Free Energy Integral}
\label{sec:freenergy}
In this section we discuss our evaluation of the integral
\begin{equation}
 f[i\tau] = -\log\left[\int D\psi\exp\left(S\right)\right],
\label{eqn:appFreeEnergy}
\end{equation} where
\begin{align}
 S = \left[\beta' L\log(R^2)/4 - \hf\int_0^L d\sigma \alpha'\beta'|\frac{\partial \psi(\sigma)}{\partial\sigma}|^2 + (i\tau + 2\mu)|\psi(\sigma)|^2 - i R^2\tau\right],
\end{align} $\beta'=\beta N$, and $\alpha' = \alpha N^{-1}$.

The free-energy, Equation \ref{eqn:appFreeEnergy}, involves a simple harmonic oscillator with a constant external force, and
we can re-write it,
\begin{equation}
 f[i\tau] = -\hf i\tau LR^2 - \beta'L\log(R^2)/4 - \ln h[i\tau].
\end{equation}
Here $h$ is the partition function for a quantum harmonic oscillator in imaginary time,
\begin{equation}
 h[i\tau] = \int D\psi \exp \left(\int_0^L\quad d\sigma -\hf m[|\partial_\sigma \psi|^2 + \omega^2|\psi|^2]\right),
\end{equation} which has the well-known solution for periodic paths in (2+1)-D where
we have integrated the end-points over the whole plane as well,
\begin{equation}
 h[i\tau] = \frac{ e^{-\omega L} } { \left(e^{-\omega L} - 1\right)^2 },
\end{equation} where $m=\alpha'\beta'$ and $\omega^2 = (i\tau + 2\mu)/(\alpha'\beta')$ (\cite{Brown:1992},\cite{Zee:2003}).

Let us make a change of variables $\lambda = i\tau + 2\mu$.  Then the free-energy reads
\begin{equation}
 f[\lambda] = (\mu - \hf\lambda)LR^2 - \beta' L\log(R^2)/4 - \ln\frac{e^{-\omega L }}{\left(e^{-\omega L} - 1\right)^2},
\end{equation} 
where $\omega = \sqrt{\lambda/(\alpha'\beta')}$.

\centering
\large
{\bf Acknowledgments}
\flushleft
\normalsize
This work is supported by ARO grant W911NF-05-1-0001 and DOE grant 
DE-FG02-04ER25616.
\pagebreak
\bibliography{pimcposter}

\begin{thebibliography}{20}
\providecommand{\natexlab}[1]{#1}
\providecommand{\url}[1]{\texttt{#1}}
\expandafter\ifx\csname urlstyle\endcsname\relax
  \providecommand{\doi}[1]{doi: #1}\else
  \providecommand{\doi}{doi: \begingroup \urlstyle{rm}\Url}\fi

\bibitem[{Abrikosov}(1957)]{Abrikosov:1957}
A.~A. {Abrikosov}.
\newblock On the magnetic properties of superconductors of the second group.
\newblock \emph{Soviet Physics JETP}, 5\penalty0 (6):\penalty0 1442--52, 1957.

\bibitem[{Andersen} and {Lim}(2006)]{Andersen2:2006}
T.~D. {Andersen} and C.~C. {Lim}.
\newblock Statistical equilibrium of trapped slender vortex filaments.
\newblock Submitted to JFM, September 2006.

\bibitem[Berdichevsky(1998)]{Berdichevsky:1998}
V.~Berdichevsky.
\newblock Statistical mechanics of vortex lines.
\newblock 57:\penalty0 2885, 1998.

\bibitem[{Berlin} and {Kac}(1952)]{Berlin:1952}
T.~H. {Berlin} and M.~{Kac}.
\newblock The spherical model of a ferromagnet.
\newblock \emph{Phys. Rev.}, 86\penalty0 (6):\penalty0 821, 1952.

\bibitem[{Brown}(1992)]{Brown:1992}
Lowell~S. {Brown}.
\newblock \emph{Quantum Field Theory}.
\newblock Cambridge UP, Cambridge, 1992.

\bibitem[{Ceperley}(1995)]{Ceperley:1995}
D.~M. {Ceperley}.
\newblock Path integrals in the theory of condensed helium.
\newblock \emph{Rev. o. Mod. Phys.}, 67:\penalty0 279, April 1995.

\bibitem[{Feynman} and {Wheeler}(1948)]{Feynman:1948}
R.~P. {Feynman} and J.~W. {Wheeler}.
\newblock Space-time approach to non-relativistic quantum mechanics.
\newblock \emph{Rev. o. Mod. Phys.}, 20:\penalty0 367, 1948.

\bibitem[{Flandoli} and {Gubinelli}(2002)]{Flandoli:2002}
F.~{Flandoli} and M.~{Gubinelli}.
\newblock The gibbs ensemble of a vortex filament.
\newblock \emph{Prob. Theory \& Rel. Fields}, 122\penalty0 (2):\penalty0 317,
  2002.

\bibitem[{Hartman} and {Weichman}(1995)]{Hartman:1995}
J.~W. {Hartman} and P.~B. {Weichman}.
\newblock The spherical model for a quantum spin glass.
\newblock \emph{Phys. Rev. Lett.}, 74\penalty0 (23):\penalty0 4584, 1995.

\bibitem[{Kevlahan}(2005)]{Kevlahan:2005}
N.~K.-R {Kevlahan}.
\newblock Stochastic differential equation models of vortex merging and
  reconnection.
\newblock \emph{Phys. of Fluids}, 17:\penalty0 065107, 2005.

\bibitem[{Klein} et~al.(1995){Klein}, {Majda}, and {Damodaran}]{Klein:1995}
R.~{Klein}, A.~{Majda}, and K.~{Damodaran}.
\newblock Simplified equation for the interaction of nearly parallel vortex
  filaments.
\newblock \emph{J. Fluid Mech.}, 288:\penalty0 201--48, 1995.

\bibitem[{Lim}(2006)]{Lim1:2006}
C.~C. {Lim}.
\newblock Phase transitions to super-rotation in a coupled barotropic fluid -
  rotating sphere system.
\newblock Moscow, Aug 2006. IUTAM, Springer-Verlag.

\bibitem[Lim and Assad(2005)]{Lim:2005}
C.~C. Lim and S.~M. Assad.
\newblock Self-containment radius for rotating planar flows, single-signed
  vortex gas and electron plasma.
\newblock \emph{R \& C Dynamics}, 10:\penalty0 240--54, 2005.

\bibitem[{Lim} and {Nebus}(2006)]{Lim2:2006}
C.~C. {Lim} and J.~{Nebus}.
\newblock \emph{Vorticity Statistical Mechanics and Monte-Carlo Simulations}.
\newblock Springer, New York, 2006.

\bibitem[{Lions} and {Majda}(2000)]{Lions:2000}
P-L. {Lions} and A.~J. {Majda}.
\newblock Equilibrium statistical theory for nearly parallel vortex filaments.
\newblock pages 76--142. CPAM, 2000.

\bibitem[{Nordborg} and {Blatter}(1998)]{Nordborg:1998}
H.~{Nordborg} and G.~{Blatter}.
\newblock Numerical study of vortex matter using the bose model: First-order
  melting and entanglement.
\newblock \emph{Phys. Rev. B}, 58\penalty0 (21):\penalty0 14556, 1998.

\bibitem[{Onsager}(1949)]{Onsager:1949}
L.~{Onsager}.
\newblock Statistical hydrodynamics.
\newblock \emph{Nuovo Cimento Suppl.}, 6:\penalty0 279--87, 1949.

\bibitem[{Sen} et~al.(2001){Sen}, {Trivedi}, and {Ceperley}]{Sen:2001}
P.~{Sen}, N.~{Trivedi}, and D.~M. {Ceperley}.
\newblock Simulation of flux lines with columnar pins: bose glass and entangled
  liquids.
\newblock \emph{Phys. Rev. Lett.}, 86\penalty0 (18):\penalty0 4092, 2001.

\bibitem[{Tsubota} et~al.(2003){Tsubota}, {Araki}, and
  {Barenghi}]{Tsubota:2003}
M.~{Tsubota}, T.~{Araki}, and C.~F. {Barenghi}.
\newblock Rotating superfluid turbulence.
\newblock \emph{Phys. Rev. Lett.}, 90:\penalty0 205301, 2003.

\bibitem[{Zee}(2003)]{Zee:2003}
A.~{Zee}.
\newblock \emph{Quantum Field Theory in a Nutshell}.
\newblock Princeton UP, Princeton, 2003.

\end{thebibliography}
\end{document}